\def\beam{{\mathrm{b}}}

\def\diff{\mathrm{d}}
\def\si{\sigma_{\mathrm{I}}}
\def\vvec{\boldsymbol v}

\def\vpar{v_{\|}}
\def\vper{v_{\bot}}

\documentclass[epsfig,amsmath,amssymb]{elsart}
\usepackage{amsmath}
\usepackage{amssymb}
\usepackage{epsfig}
\usepackage{graphicx}
\usepackage{epic}
\usepackage{eepic}
\usepackage{dcolumn}
\usepackage{bm}

\begin{document}
\begin{frontmatter}

\title{Crossover from a fission-evaporation scenario towards
multifragmentation in spallation reactions}

\author[1,2]{P. Napolitani}

\address[1]{GANIL (DSM-CEA/IN2P3-CNRS), Blvd. H. Becquerel, 14076 Caen, France}
\address[2]{LPC (IN2P3-CNRS), 14076 Caen, France}

\begin{abstract}
	Mostly for the purpose of applications for the energy and 
the environment and for the design of sources of neutrons or exotic 
nuclides, intense research has been dedicated to spallation, 
induced by protons or light projectiles at incident energies of 
around 1 GeV.
	In this energy range, while multifragmentation has still a 
minor share in the total reaction cross section, it was observed to 
have, together with fission, a prominent role in the production and 
the kinematics of intermediate-mass fragments, so as to condition 
the whole production of light and heavy nuclides.
	The experimental observables we dispose of attribute rather
elusive properties to the intermediate-mass fragments and do not
allow to classify them within one exclusive picture which is either 
multifragmentation or fission.
	Indeed, these two decay mechanisms, driven by different 
kinds of instabilities, exhibit behaviours which are closely 
comparable.
	High-resolution measurements of the reaction kinematics 
trace the way for probing finer features of the reaction 
kinematics.
\vspace{0.5cm}
\newline 
{\em 
Conference proceedings: International Meeting
``Selected topics on nuclear methods for non-nuclear
applications'', September 27-30, 2006, Varna}
\end{abstract}
\end{frontmatter}


\thispagestyle{empty}
\section[]{Introduction \label{sect1}}
	Sixty years passed from Serber's early
description~\cite{Serber47} of high-energy nuclear reactions 
induced by nucleons and light nuclei.
	In their main outline, such reactions are describes as the 
exciting of an atomic nucleus,
followed by a decay process in several nuclides, 
clusters, protons and neutrons~\cite{Hufner85}.
	Several different nuclear systems have 
been explored with beams of various incident energies, at different 
facilities and with different experimental techniques.
	Numerous experimental results and various observables on 
the production of residues, on the kinematics of the emission of 
ejectiles and fragments, and on the correlations inspired and 
constrained the physical models.

	At incident energies of few hundred MeV per nucleon, an
excited and fully equilibrated complex, named compound nucleus, is
formed and successively de-excites by mainly fission-evaporation 
decays.
	When the excitation energy of the hot nucleus exceeds the 
threshold for emission of particles or clusters (including 
fission), the system has the possibility to decay by any of the 
open channels.
	If the excited system is not too hot, the favoured process 
is the reordering of its configurations: a great number of 
arrangements are available where all nucleons remain in states 
below the continuum, occupying excited single-particle levels
around the Fermi surface.
	Rather seldom, compared with this thermal chaotic motion of 
the system, one nucleon acquires enough energy to pass above the 
continuum and may eventually leave the nucleus.
	This picture was extended to include the production of 
intermediate-mass fragments by cluster decay and oscillations in 
fission direction as well.
	In this respect, there would be a gradual transition from 
very asymmetric to symmetric configurations in the binary split of 
the decaying compound nucleus, so that evaporation of nucleons and 
light nuclei on the one hand and symmetric fission on the other 
hand are just the opposite extremes of the manifestation of the 
same process.
	This generalisation, introduced by 
Moretto~\cite{Moretto75,Moretto89}, allows to name fission in a 
generalised sense all (binary) decays of a compound nucleus.
	Since this decay is a rare process, one evaporation event, 
or fission event, proceeds after the other, sequentially.

	At incident energies of some GeV per nucleon, a very highly
excited composite nuclear system is formed; it undergoes a violent 
de-excitation processes, named multifragmentation, which manifests 
by the production of several massive fragments filling broad 
kinetic-energy spectra.
	General reviews on this process can be found in 
refs.~\cite{Moretto93,Bonsignori01}; a more specific review 
treating multifragmentation induced with high-energy proton beams
can be found in ref.~\cite{Viola06}.
	The amount of experimental evidence suggests to describe
the de-excitation as a simultaneous disassembling of the hot 
nucleus in several constituents.
	From the experimental observation of angle correlations 
between fragments, it was found that the emission of two fragments
with small relative angles is highly improbable.
	Such a dependence on relative emission angles is on the 
contrary absent for the ejectiles of a sequential evaporation 
process~\cite{Karnaukhov03a}.
	The disintegration is so rapid to be considered
``simultaneous'', in the sense that it evolves in so short a time 
interval ($10^{-22}$-$10^{-21}$s) that fragments can still 
exchange interactions while they are accelerated in their mutual
Coulomb field.
	On the other hand, the process is expected to be still 
sufficiently ``slow'' to exceed the relaxation time of the strong 
interactions and, for this reason, the system is assumed to be 
thermalised before it disintegrates.
	Within a thermodynamical picture, multifragmentation
is a phenomenon related to the equation of state of nuclear 
matter~\cite{Bertsch83,Siemens83}; with a certain resemblance with 
Van-der-Waals fluid, hot bulk matter would enter the liquid-gas 
coexistence region of the phase diagram and separate in the 
corresponding coexisting liquid and gas phases (see 
refs.~\cite{Bonasera00,Richert01} for review), driven by
a rapid amplification of spinodal instabilities~\cite{Chomaz04}.
	According to the same picture, when the composite system 
reaches the conditions for disintegrating, it should be diluted to
some fractions of the nuclear saturation density.
	These low densities are reached as a result of a dynamical
process of expansion, which is expected to explain the high 
velocity of the fragments observed experimentally.
	There are however also alternative mechanisms proposed to 
explain the experimental features of the multifragmentation process 
which are close to the fully equilibrated compound-nucleus 
decay~\cite{Friedman83,Lopez89}.

	In recent years, mostly for the purpose of 
energetic and environmental applications, as well as for the 
production of beams of neutrons or exotic nuclides in accelerator 
facilities, increasing interest is devoted to spallation 
reactions induced by protons or deuterons at incident energies 
close to 1 GeV.
	At this energy, the reaction is situated somehow in between 
the two scenarios described above and is 
particularly interesting for studying the transition from fission 
to multifragmentation.
	The physics underlying the spallation process in this 
energy range is however rather uncontrolled.
	From the experimental side, the approaches used up to
now have always provided a partial survey of all the observables
which are necessary to formulate conclusive answers.
	Inclusive approaches are best suited for the 
identification of the nuclides, for the measurement of the 
corresponding production cross sections and kinetic-energy 
distributions; they however neither provide particle correlations, 
like the multiplicity of the fragments and particles formed in the 
reaction, nor angle (or velocity) correlations, necessary to 
probe the reaction kinematics for each single event.
	Correlation observables are the specificity of 
exclusive approaches, based on the employment of multidetectors
which, at the moment, are still not able to provide the full
isotopic identification up to heavy elements and high-resolution
velocity measurements.
	On the other hand, from the modelling side, several 
different strategies based either
on a generalised fission-evaporation scenario~\cite{Moretto75,Charity88} 
or on a statistical description of 
multifragmentation~\cite{Randrup81,Fai82,Randrup93,Botvina85,Bondorf85,Bondorf95,Gross90,Gross97} 
have been able to describe the same experimental data with 
comparable accuracy.

\section{Residue production in spallation induced by 1 GeV protons \label{sect2}}
\begin{figure}[b!]\begin{center}
\includegraphics[angle=0, width=0.8\columnwidth]{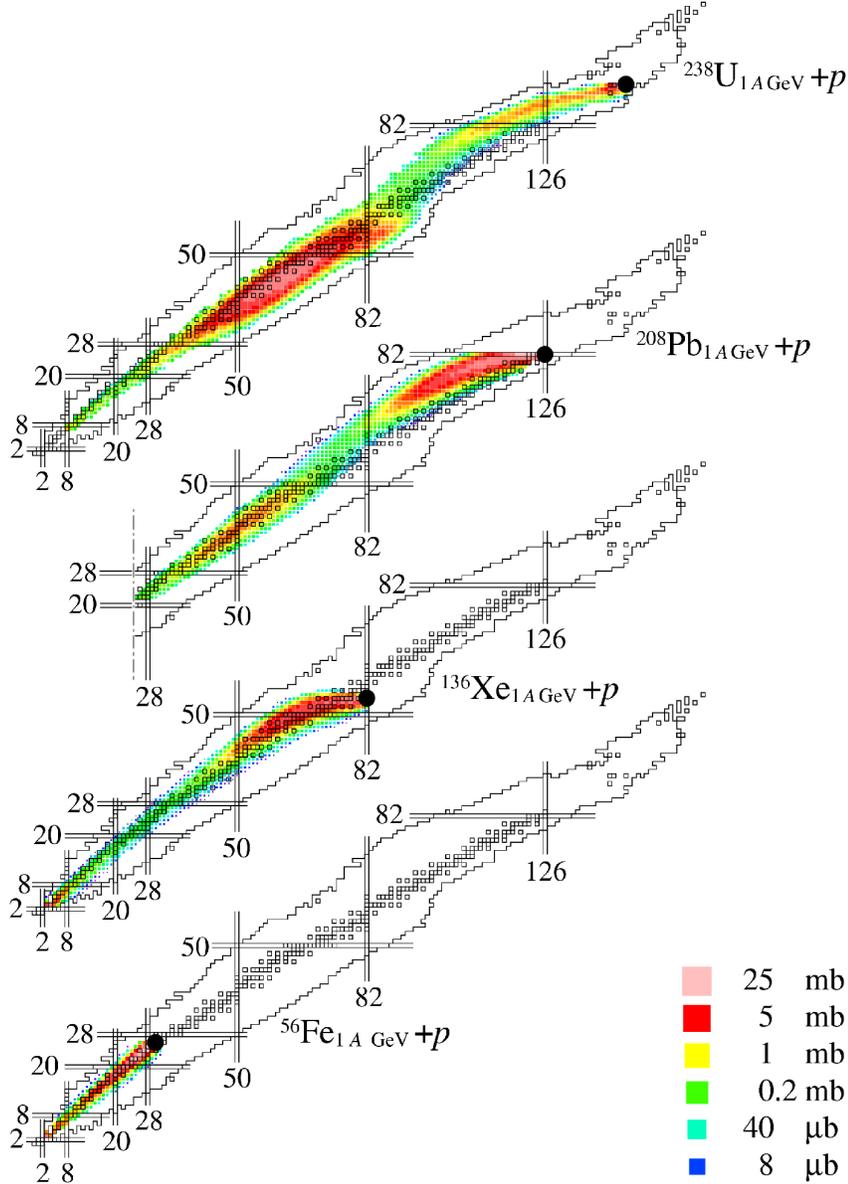}
\end{center}\caption
{
	Experimental survey on the distribution of spallation 
residues presented in nuclide charts for the systems
$^{238}$U$_{(1\,A\,\textrm{GeV})}+p$~\cite{Armbruster04,Taieb03,Bernas03,Bernas06,Ricciardi06},
$^{208}$Pb$_{(1\,A\,\textrm{GeV})}+p$~\cite{Enqvist01,Kelic04},
$^{136}$Xe$_{(1\,A\,\textrm{GeV})}+p$~\cite{Napolitani06}, and
$^{56}$Fe$_{(1\,A\,\textrm{GeV})}+p$~\cite{Napolitani04,Villagrasa?}.
Filled circles indicate the projectiles (in inverse kinematics).
}
\label{fig1}
\end{figure}

	The measurement of the production of spallation residues
at relativistic energy has been the purpose of several years of 
experiments performed at the FRagment 
Separator~\cite{Geissel92,Schmidt87} (GSI, Darmstadt).
	The reactions were measured in inverse kinematics, by
directing heavy-ion beams at various energies on a target of
liquid hydrogen or deuterium.
	Fig.~\ref{fig1} presents a survey on the nuclide
production for the  spallation reactions induced by protons
at 1 GeV which were measured at the FRagment Separator.
%
%

	For the reaction $^{238}$U$_{(1\,A\,\textrm{GeV})}
+p$~\cite{Armbruster04} the nuclide distribution reflects the 
interplay of more decay patterns.
	Quite independently on the entrance channel (and in 
particular the neutron enrichment of the compound nucleus), 
evaporation residues~\cite{Taieb03} populate the neutron-poor side 
of the nuclide chart around the residue corridor~\cite{Dufour82}, 
which corresponds to the situation where~\cite{Charity98} 
$\mathrm{d}N/\mathrm{d}Z=\langle\Gamma_{N}/\Gamma_{Z}\rangle$,
where $\Gamma_{Z}$ and $\Gamma_{N}$ are the proton- and 
neutron-emission width, respectively.
 	This is for instance the reason why neutron-rich nuclei
are favoured materials for neutron sources.
	In the centre of the nuclide distribution an imposing
high-energy fission hump emerges between two shoulders generated by 
low-energy fission channels~\cite{Bernas03}, which are exotic and 
asymmetric.
	Other asymmetric fission channels are determined 
by high-energy fission, as an extension of the central hump; they
overlap with the evaporation production on one side~\cite{Bernas06} 
and extend to the intermediate-mass fragments~\cite{Ricciardi06} on 
the opposite side.

	While uranium is an attractive material for the production 
of exotic nuclei by low-energy fission, lead is preferred as a 
neutron source because, as the measurement of 
$^{208}$Pb$_{(1\,A\,\textrm{GeV})}+p$ 
reveals~\cite{Enqvist01,Kelic04}, fission is much reduced with 
respect to evaporation channels.

	Iron was measured because it is a structural material in 
nuclear installations.
	From a phenomenology point of view, the results on the 
reaction $^{56}$Fe$_{(1\,A\,\textrm{GeV})}+p$~\cite{Napolitani04,Villagrasa?} brought the 
attention to light systems, below the Businaro-Gallone
point~\cite{Businaro55a,Businaro55b}.
	At variance with heavy systems, the fission potential 
becomes convex and fission channels are asymmetric.
	Moreover, the excitation energy is larger in lighter systems
when reactions are compared at the same incident energy.
	As a result of reaction models~\cite{Napolitani04}, for 
small impact parameters the excitation energy of the composite 
system generated in the reaction 
$^{56}$Fe$_{(1\,A\,\textrm{GeV})}+p$ is found to even exceed
$3$ MeV per nucleon; around this value multifragmentation is 
expected to set in.
	For the production of intermediate-mass fragments,
this process sums up to the asymmetric fission channels.

	The measurement of $^{136}$Xe$_{(1\,A\,\textrm{GeV})}+p$ 
was performed on purpose to analyse the intermediate-mass-fragment
production and its influence on the overall decay mechanism.
	In this respect, $^{136}$Xe offers an optimum point of 
observation: symmetric fission, which in heavier systems hides the
intermediate-mass-fragment production, is suppressed because the 
system is slightly below the Businaro-Gallone point; then, below 
the Businaro-Gallone point, $^{136}$Xe is the stable nuclide with the 
largest neutron excess $N-Z$, and it approaches the neutron 
enrichment $N/Z$ of $^{208}$Pb, so that a comparison can be made 
between the two systems; and finally, the system
$^{136}$Xe$_{(1\,A\,\textrm{GeV})}+p$ is excited right enough to 
still allow for some oscillations towards multifragmentation.
	The results on the nuclide production~\cite{Napolitani06}
illustrate that, besides a close similarity with the system
$^{208}$Pb$_{(1\,A\,\textrm{GeV})}+p$ for the evaporation 
features, beyond a mass loss of around $\Delta A = 70$, the 
ridge of the residue production abandons the neutron-poor side 
of the nuclide chart around $Z=40$ and migrates progressively 
towards the neutron-rich side for lighter residues.
	The lightest residues even populate the neutron rich side 
of the nuclide chart with respect to the valley of stability.

	It may be remarked that both asymmetric fission channels and
multifragmentation of a neutron-rich composite system produce 
residues with high neutron excess in the average; the successive 
evaporation of residues, which is expected due to the high mean 
excitation energy of the system, contributes to further dissipate  
the traces of the initial stage of the decay process~\cite{HenzlovaPHD}.
	Qualitatively, both the two mechanisms also result in a 
similar U-shape of the mass distribution of residues, with 
decreasing depth of the hollow for increasing mean excitation 
energy of the system.
	For this reason, the residue production alone is a rather 
elusive observable and it could be reproduced with comparable 
quality both on the basis of multifragmentation processes and 
asymmetric fission.


\section{Kinematics \label{sect3}}
	More robust observables for the reaction mechanism are 
those related to the Coulomb field experienced by the fragments.
	In exclusive experiments, these are angle correlations 
among fragments, which reflect the role of the Coulomb field in 
each event; such experimental strategy, based on multidetectors and 
largely used for ion-ion collisions or for multifragmentation 
induced by high-energy protons, was unfortunately not employed in 
the study of spallation in the incident-energy range of around 
1 GeV.
	In inclusive experiments, the observables probing the 
Coulomb field are the mean 
quantities related to the momentum distribution of fragments.
	Evidently, inclusive observables mixes up the 
contributions of all processes responsible for the formation of one 
given fragment; therefore, the identification of the reaction 
process can not be imposed from a selection of separate observables
(like the fragment multiplicity or the transfered energy of light 
charged particle in exclusive experiments~\cite{Pichon06}), 
but it should be extracted by the momentum distribution itself.
	Up to a certain extent, this is possible with magnetic 
spectrometers operated in inverse kinematics, as
in the case of the experiments discussed in 
section~\ref{sect2}, where the high-resolution momentum 
distributions could be deduced from the measurement of the magnetic 
rigidity $B\rho$, as precise as $5 \cdot 10^{-4}$ (FWHM)
for individual reaction products.
%
%
\begin{figure}[b!]\begin{center}
\includegraphics[angle=0, width=1\columnwidth]{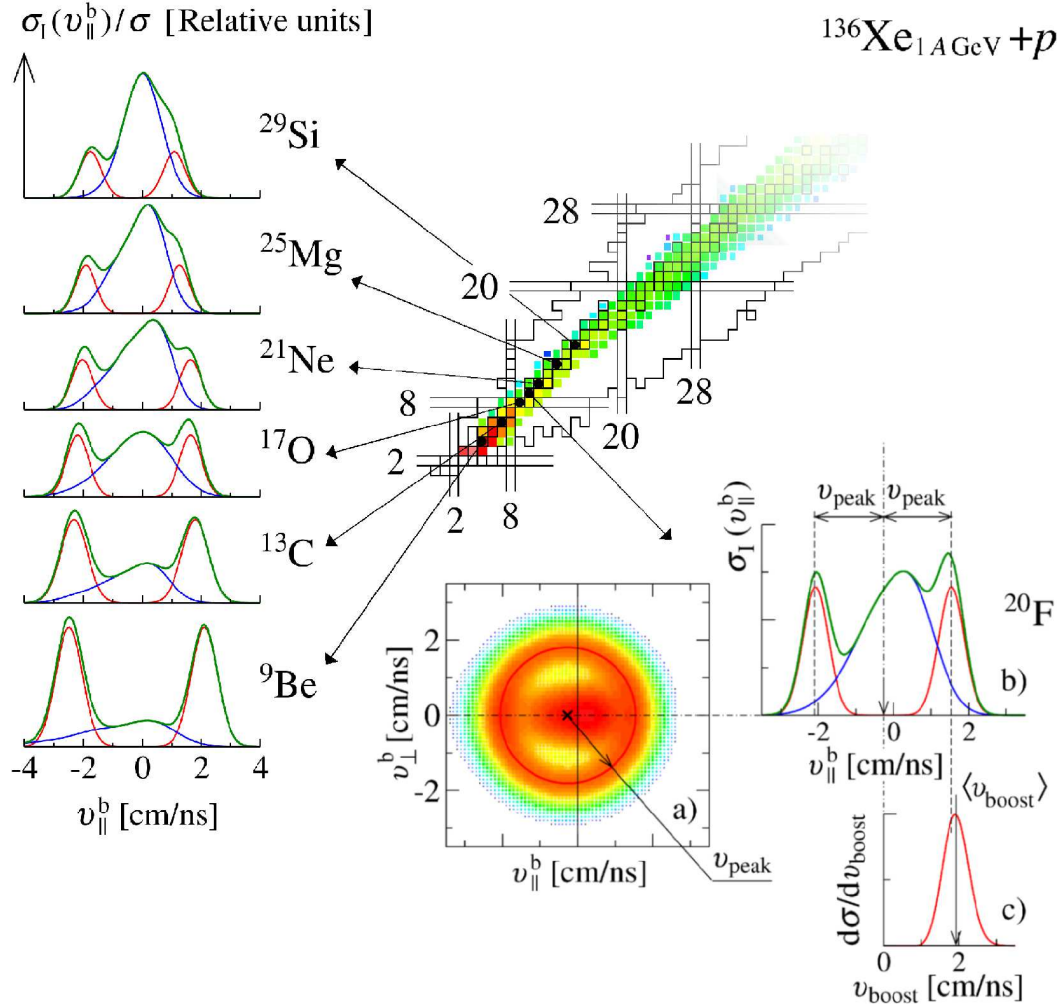}
\end{center}\caption
{
	Distribution of invariant cross section for some 
intermediate mass fragments produced in the reaction 
$^{136}$Xe$_{(1\,A\,\textrm{GeV})}+p$~\cite{Napolitani06}.
	The corresponding nuclides are indicated on a portion of
the fragment distribution, on the nuclide chart.
	Two kinematical modes, one convex in the centre, the other
concave in the centre, are indicated.
	For the nuclide $^{20}$F more details are shown:
(a) The planar cut along the beam axis of the reconstructed full 
distribution $\diff\sigma/\diff\vvec^\beam$ in the beam 
frame as a cluster plot. 
	The concave mode is indicated by the corresponding circular 
ridge of radius $v_{\mathrm{peak}}$. 
(b) The invariant-cross-section distribution, as for the other 
nuclides.
(C)	Reconstructed cross-section distribution for the 
	concave mode as a function of the boost
	velocity in the source frame $v_{\mathrm{boost}}$.
	The plot allows to calculate the mean boost 
	$\langle v_{\mathrm{boost}}\rangle$, which differs from 
	$v_{\mathrm{peak}}$. 
}
\label{fig2}
\end{figure}

	A dedicated analysis procedure 
(refs.~\cite{Napolitani04,Napolitani06} for detailed discussion)
was applied to reconstruct invariant cross sections from the 
inclusive measurement of the momentum distribution at the FRagment
Separator.
	An overview of the result is shown in fig.~\ref{fig2} for 
some light resides of the reaction $^{136}$Xe$_{(1\,A\,\textrm{GeV})}+p$.
	More details of this observable are shown for the nuclide
$^{20}$F: in the insert (a), a planar cut along the beam axis of 
the full velocity distribution in the projectile frame is shown in
the $\vper \times \vpar$ space ($\vper$ and $\vpar$ are 
respectively the perpendicular and parallel velocity components of 
the fragment $^{20}$F in the beam frame), as reconstructed from the 
measured momentum distribution; by selecting the velocities 
aligned along the beam axis, this representation can be reduced to
the distribution of invariant cross section $\si$ as a function of
$\vpar$ in the projectile frame, shown in the insert (b).
	The distribution of invariant cross section $\si$ of all
the intermediate mas fragments formed in the reaction 
$^{136}$Xe$_{(1\,A\,\textrm{GeV})}+p$ results from the overlap of 
two shapes: a component with a convex centre and a component with a 
concave centre.

	A ``convex'' mode describes two completely different 
situations.
	In a first case, it describes the velocity distribution of 
evaporation residues after sequential emission of nucleons and 
clusters.
	At these incident energies, the most probable excitation 
energy is just sufficient for the emission of few nucleons and the
probability of higher excitations, connected with longer 
evaporation paths, can only decrease progressively;
this behaviour ensures that the distribution of (heaviest) 
evaporation residues drops in cross section monotonically when 
moving away from the projectile and its contribution to the 
production of intermediate mas fragments is invisible for all the 
systems described in fig.~\ref{fig1}.
	For these systems, a convex distribution of invariant cross 
section associated to intermediate mass fragments rather indicates 
a multifragmentation process; in particular, the convex mode probes 
the Coulomb field produced by the disassembling of the system in 
more fragments having a comparable size.
	The explanation for the asymmetry of this mode, 
characterised by a long tail in backward direction and a maximum 
shifted forward, may be searched in possible dynamical effets.
	It may be interesting to remark that the convex mode,
associated to multifragmentation, can not be observed in direct 
kinematics because a threshold in kinetic energy hides the centre 
of the distribution of invariant cross section.
	The ``concave mode'' reflects a
strong Coulomb boost as experienced by fission fragments.
	When this pattern is associated to intermediate-mass
fragments, it may indicate an asymmetric fission process as well as
the possible extension of multifragmentation events to very 
asymmetric partitioning configuration; in this latter case, the 
kinematics is determined by the repulsion exerced by 
one very heavy fragment and it should 
not be disturbed greatly by the presence of more than one light 
fragment.
	In conclusion, a concave component in the distribution of
invariant cross section of intermediate-mass fragments indicates 
that the system certainly broke up asymmetrically, but the 
attribution of the process to fission or multifragmentation stays
uncertain on the basis of this inclusive observable.

%
%
\begin{figure}[b!]\begin{center}
\includegraphics[angle=0, width=1\columnwidth]{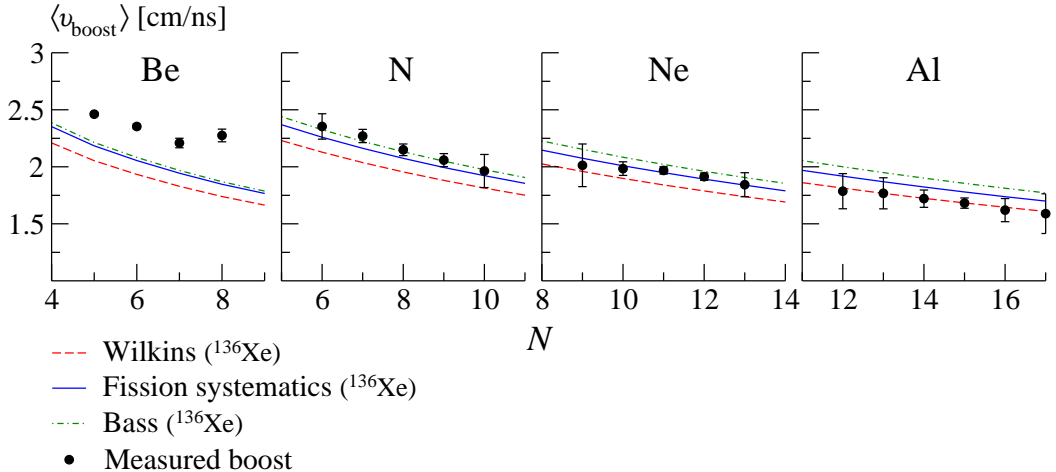}
\end{center}\caption
{
	Evolution of the reconstructed mean boost, deduced for the 
convex kinetic mode as a function of the neutron number for some
light elements in the reaction 
$^{136}$Xe$_{(1\,A\,\textrm{GeV})}+p$~\cite{Napolitani06}.
	The measurement is compared with three expectations for the
Coulomb boost in the split of $^{136}$Xe, according to the
total-kinetic-energy systematics of Tavares and 
Terranova~\cite{Tavares92}, the scission-point model of Wilkins et 
al.~\cite{Wilkins76, Bockstiegel97} and the nucleus-nucleus fusion
model of Bass~\cite{Bass79, Bass80}.
}
\label{fig3}
\end{figure}
	In fig.~\ref{fig2}, the insert (c) represents for the 
nuclide $^{20}$F the convex component alone in polar coordinates in
the frame of the corresponding average emitting source.
	The mean value of this spectrum 
$\langle v_{\mathrm{boost}}\rangle$ would 
be consistent with a fission barrier if the process were 
exclusively reduced to fission.
	Such test is shown in fig.~\ref{fig3} for some light 
elements produced in the system
$^{136}$Xe$_{(1\,A\,\textrm{GeV})}+p$.
	The figure compares the measured mean boost 
$\langle v_{\mathrm{boost}}\rangle$ with the highest expected 
value, determined by the split of the heaviest possible system, 
$^{136}$Xe.
	The expected boost is the highest when the fission 
fragments are considered not deformed and subjected to the fusion 
potential, as described by the empirical formula of 
Bass~\cite{Bass79, Bass80}.
	The lowest value is calculated when the fission fragments 
are considered deformed and joined through a neck according to the 
liquid-drop picture (scission-point model of Wilkins et 
al.~\cite{Wilkins76, Bockstiegel97}).
	In between, is situated the value calculated from a 
systematics of total kinetic energy for light fissioning 
nuclei~\cite{Tavares92}, further modified in 
ref.~\cite{Napolitani04} to describe asymmetric splits.
	The lighter are the elements, the more all the three 
expectations underpredict the data.
	This test suggests that the concave mode should be 
alimented, in addition to fission, also by multifragmentation 
channels which could be related to an expansion process and could
enhance the mean boost.

	Similar experimental observations were collected for the 
system $^{56}$Fe$_{(1\,A\,\textrm{GeV})}+p$~\cite{Napolitani04}.
	In the system $^{238}$U$_{(1\,A\,\textrm{GeV})}+p$, the
intermediate-mass fragments were fully described by the concave 
mode; also in this case, the corresponding mean boost was 
characterised by very high values~\cite{Ricciardi06} so that it is 
tempting to make a comparison with the lighter systems 
$^{56}$Fe$_{(1\,A\,\textrm{GeV})}+p$ and
$^{136}$Xe$_{(1\,A\,\textrm{GeV})}+p$.

\section{Concluding remarks \label{sect4}}
	In conclusion, the coupling of kinematical and production
observables is a rather robust approach to study the reaction 
mechanism in such a complex situation as spallation reactions at
incident energies of around 1 GeV.
	This approach indicated that multifragmentation is a 
possible decay pattern and determines the kinematics by producing a 
convex component in the distribution of invariant cross section.
	More difficult is to define its possible extension to the
asymmetric breakups of the system, determined by a concave 
component in the distribution of invariant cross section. 
Is the observation of a very high mean boost in the asymmetric 
splits a signature of dynamical effects, like expansion, which goes 
mostly in the direction of multifragmentation?

%
%
\begin{figure}[t!]\begin{center}
\includegraphics[angle=0, width=1\columnwidth]{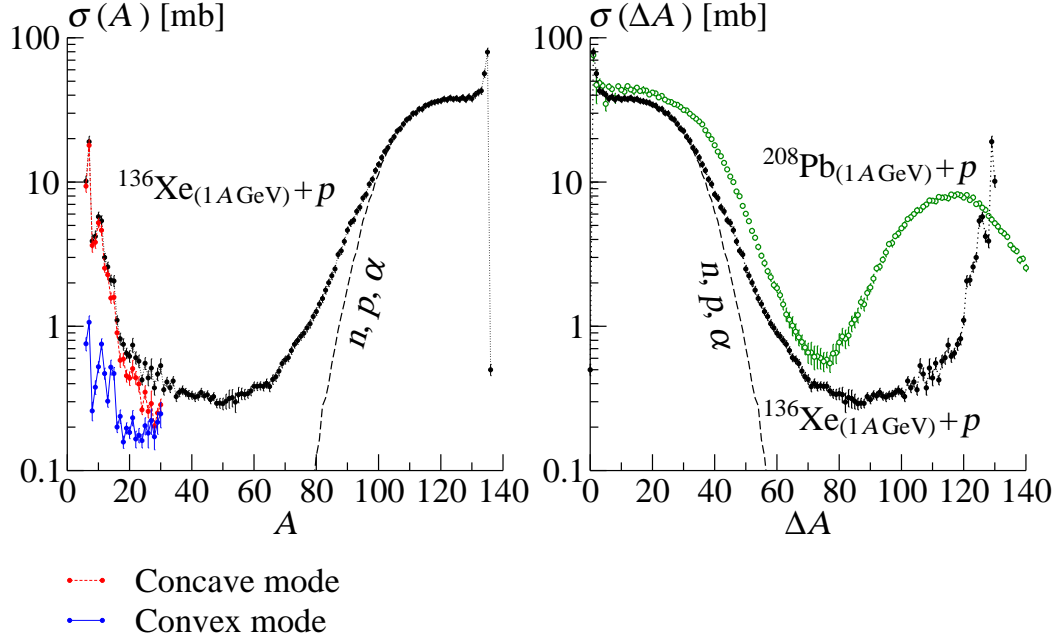}
\end{center}\caption
{
	Left panel. 
	Projections of the residue-production cross sections along 
	the mass number for the system
	$^{136}$Xe$_{(1\,A\,\textrm{GeV})}+p$~\cite{Napolitani06}.
	For intermediate mass fragments the 
	contributions from the two kinematical modes, concave and 
	convex, are indicated.
	A Weisskopf calculation for the expected contribution of 
	the evaporation of protons, neutrons and alpha particles 
	is shown. 
	Right panel. 
	Mass-loss distribution of production cross sections.
	The experimental results obtained for the systems
	$^{136}$Xe$_{(1\,A\,\textrm{GeV})}+p$~\cite{Napolitani06} 
	and $^{208}$Pb$_{(1\,A\,\textrm{GeV})}+p$~\cite{Enqvist01}
	are compared.
	The same Weisskopf calculation for the system
	$^{136}$Xe$_{(1\,A\,\textrm{GeV})}+p$
	shown in the left panel, is repeated for comparison.
}
\label{fig4}
\end{figure}
	The left panel of fig.~\ref{fig4} shows the mass 
distribution of the residues of the system 
$^{136}$Xe$_{(1\,A\,\textrm{GeV})}+p$.
	It also shows for the concave and convex kinematical modes 
the corresponding contribution to the production cross section.
	The limitations of the experimental approach did not allow
to complete the distribution for these two components over the 
whole mass distribution.
	However, we can appreciate a rise in cross section of the 
convex mode up to joining the value of the concave mode.
	The convex mode is expected to extend further and to be the
dominant mode in the region of the hollow of the mass distribution.
	The concave mode, which gives the highest contribution in 
the production of the light nuclides, drops steeply in cross 
section in proximity of the hollow.
	Since this mode supposes the existence of a heavy partner 
in a split of mostly binary kind, the corresponding distribution of
cross section should also aliment the region of heavy masses.
	The side of the mass spectrum in the region of heavy masses
is often taken as a reference for testing the modelling of the 
initial stage of the reaction process (cascade),
due to its sensitivity to the excitation energy introduced in the 
system during the collision.
	However, already at incident energies of around 1 GeV, the 
contribution of the convex mode to the production of heavy masses 
should greatly modify the slope of the mass distribution for the
heavy-fragment side of the hollow, with respect to the function 
characterising a more simple mechanism where only protons, neutrons
and alpha particles are emitted.
	Such function, which can be estimated by a Weisskopf
calculation, drops to imperceptible cross sections for a mass loss
$\Delta A$ of less than sixty units and deviates form the 
experimental distribution already after $\Delta A\approx 40$.
	The right panel of fig.~\ref{fig4} shows that the cross
section evolves with the mass loss in a very similar way for the
system $^{136}$Xe$_{(1\,A\,\textrm{GeV})}+p$ and
$^{208}$Pb$_{(1\,A\,\textrm{GeV})}+p$ in the region of heavy 
masses.

	This feature reinforces the idea that the concave mode,
associated to a fast fission-like process, and observed in the 
velocity spectra for systems of different size, from 
$^{56}$Fe$_{(1\,A\,\textrm{GeV})}+p$ to
$^{238}$U$_{(1\,A\,\textrm{GeV})}+p$, is a rather general picture
in spallation reactions induced by 1 GeV protons.
	On the other hand, the concave mode, more evidently related 
to multifragmentation, is a characteristic of more excited system.

	The purpose of this report was to outline the main 
phenomenological features of spallation in the incident-energy 
range of 1 GeV per nucleon, without discussing any consequence on 
applications.
	It however derives naturally that the passage from the 
phenomenology to the modelling is necessary for application
purposes, in order to better describe the overall production of
nuclides and the kinematics of light fragments in spallation
reactions which are and will be largely employed in energetic and 
environmental applications.

\section*{Acknowledgments}

	This contribution uses the results of many experiments, 
performed at the FRagment Separator (GSI, Darmstadt) by the
collaboration of the following scientists: 
 P. Armbruster,
 L. Audouin, 
 A. Bacquias,
 J. Benlliure, 
 M. Bernas, 
 A. Boudard, 
 E. Casarejos, 
 J. J. Connell, 
 S. Czajkowski, 
 J.-E. Ducret, 
 T. Enqvist, 
 T. Faestermann, 
 B. Fernandez, 
 L. Ferrant, 
 J. S. George, 
 L. Giot,
 F. Hammache, 
 A. Heinz, 
 K. Helariutta, 
 V. Henzl, 
 D. Henzlova, 
 A. R. Junghans, 
 B. Jurado, 
 D. Karamanis, 
 A. Keli\'c, 
 R. Legrain, 
 S. Leray, 
 R. A. Mewaldt, 
 B. Mustapha, 
 M. F. Ordonez, 
 J. Pereira, 
 M. Pravikoff, 
 F. Rejmund, 
 M. V. Ricciardi,  
 K.-H. Schmidt,
 C. Schmitt, 
 C. St\'ephan, 
 K. S\"ummerer, 
 J. Ta\"ieb, 
 L. Tassan-Got, 
 C. Villagrasa, 
 F. Viv\`es, 
 C. Volant, 
 M. E. Wiedenbeck, 
 W. Wlazlo, 
 N. E. Yanasak, 
 and 
 O. Yordanov. 
%
%
%
%

\end{document}